\documentstyle[11pt,paspconf,epsf]{article}

\begin{document}

\title{Depletion curves of galaxy number counts behind cluster lenses}
\author{C. Mayen and G. Soucail}
\affil{Observatoire Midi-Pyr\'en\'ees, Laboratoire d'Astrophysique, 14
Avenue E. Belin, 31400 Toulouse, France}

\begin{abstract}
When the logarithmic slope of the galaxy counts is lower than 0.4 (this
is the case in all filters at high magnitude), the
magnification bias due to the lens makes the number density of objects
decrease. Consequently,
the radial distribution shows a typical depletion curve.
We present simulations of depletion curves obtained 
for a variety of different lens models and we 
show how the model parameters affect the depletion area.
\end{abstract}


\section{Introduction}
The magnification bias of number counts of background galaxies is used 
to determine cluster mass
distribution (Broadhurst et al., 1995; Taylor et al., 1998), the
redshift distribution of sources and to bring constraints on
cosmological parameters (Fort et al., 1997). All these methods show that
it is important to study the width and the shape of the depletion area.\\
We have simulated the depletion effect with four different lens models :
the singular isothermal sphere (SIS), the isothermal sphere with a core
(SIC), the power-law profile (PLP) and an elliptical potential (EP)
introduced by Schneider et al. (1992).
We also use two analytical
redshift distributions for background objects. The first one
contains two identical populations located at $z=1$ and
$z=2$ respectively. 
The second one is the one
used by Taylor et al. (1998). We adopt $h=0.5$, $\Omega_{\Lambda}=0$,  
$\Omega_0=1$ and $z_d=0.4$.

\section{Results and future work}

Fig. 1 shows the results.
The increase of the velocity dispersion $\sigma$ leads to an increase of the
minimum position of the curve (increase of the Einstein radius) and 
of the half width at half minimum of the depletion area which can be
explained by the sensitivity of the depletion to the mass in the outer
parts of the cluster which is bigger when $\sigma$ increases.
The intensity of the depletion is not affected by $\sigma$.\\
The introduction of a core radius $R_c$ in the model leads to the 
appearance of a hole near the cluster's center followed by a bump which
decreases as $R_c$ increases.\\
The increase of the slope $\alpha$ of the mass 
profile does not affect the minimum position of the depletion area but 
leads to an increase of the minimum value and a decrease of the half
width at half minimum of the curve (same reason as for the 
SIS : When $\alpha$
increases, the mass is concentrated in the inner part of the cluster and
the depletion effect is less extended to the outer regions).\\
The variation of the ellipticity $\epsilon$ of the potential gives  
rise to a stretching with
regard to the minimum position and the minimum value of the curve.

\begin{figure}
\plotfiddle{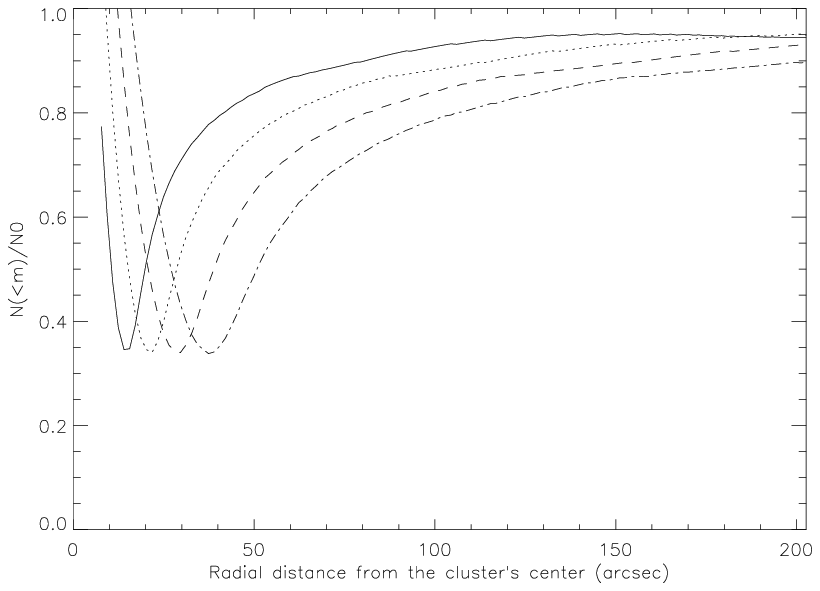}{2.0cm}{0.}{50.}{50.}{-150.}{0.}
\plotfiddle{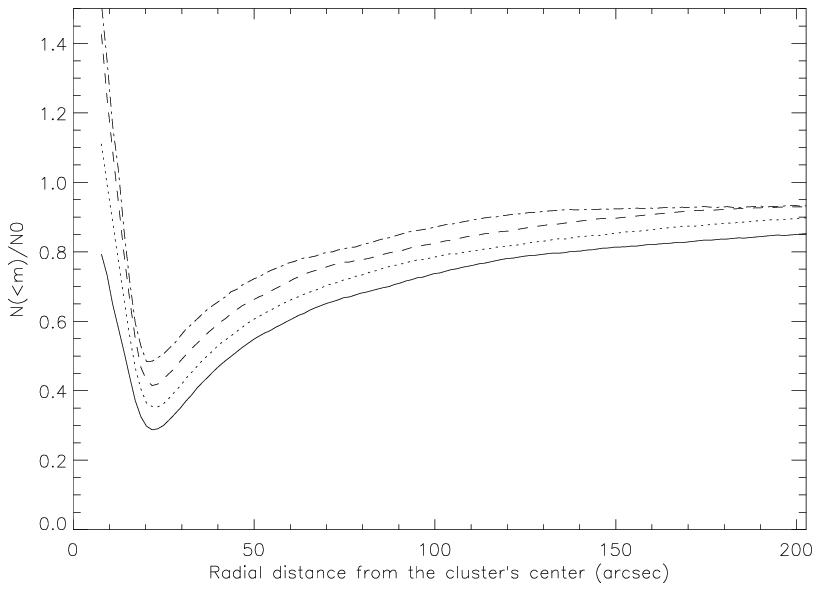}{0.0cm}{0.}{50.}{50.}{10.}{25.}
\plotfiddle{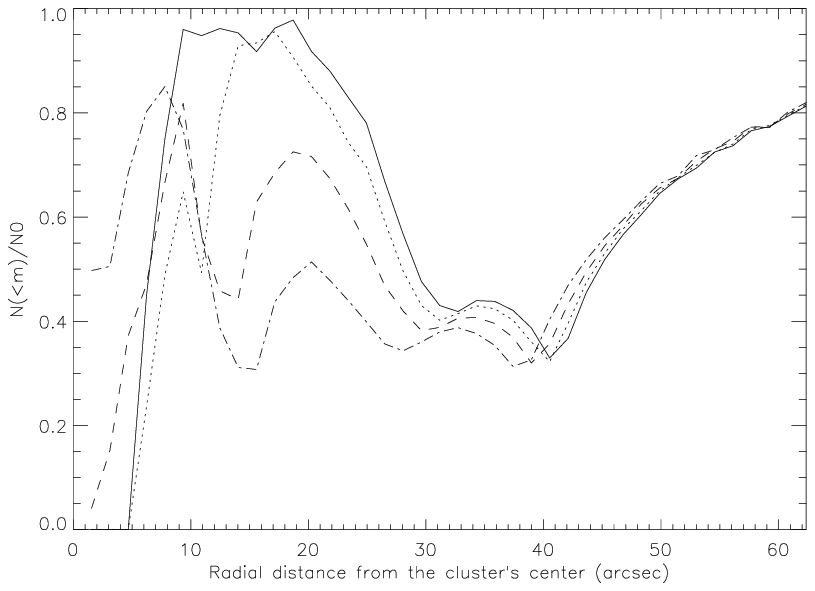}{2.0cm}{0.}{50.}{50.}{-150.}{0.}
\plotfiddle{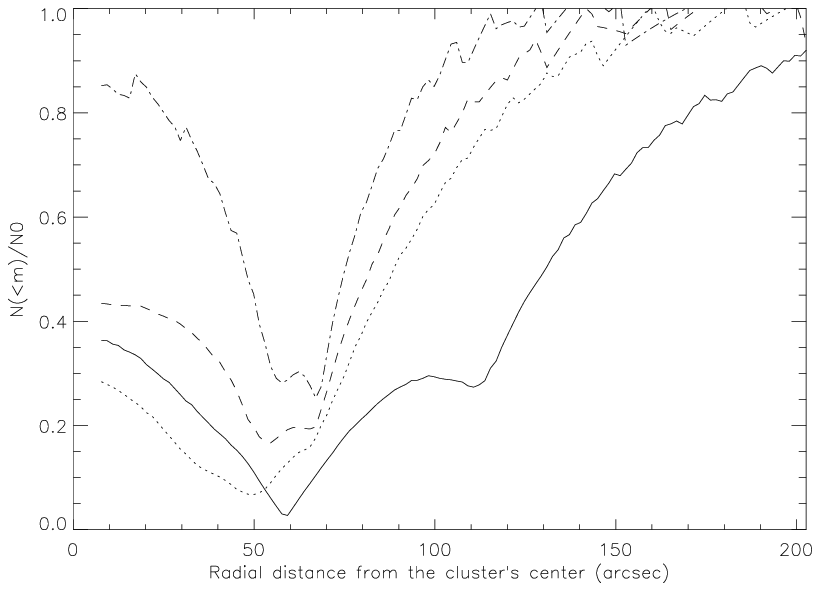}{0.0cm}{0.}{50.}{50.}{10.}{25.}
\caption{Depletion curves obtained with a SIS (top left) for $\sigma =$
1000 km/s (---), 1200 km/s (...), 1400 km/s
(\_ \_) and 1600 km/s (\_.\_); with PLP
(top right) for $\alpha =$1.8 (---), 1.9 (...), 2.0
(\_ \_) which is equivalent to the SIS and 2.1 (\_.\_); with a SIC 
(bottom left) for $R_c=$ 25 kpc (---), 50 kpc (...),  
75 kpc (\_ \_) and 100 kpc (\_.\_); with an
EP (bottom right) for $\epsilon =$0 (---),
0.3 (...), 0.4 (\_ \_) and 0.6 (\_.\_).}
\label{fig1}
\end{figure}

We have simulated the various influences of model parameters
on the typical features of the depletion curves. As these ones are 
poorly known, their influence on the
depletion area must be explored in more details before application to
real data. The effects of the filters, redshift distribution of the 
background sources and color selection on the depletion area 
have also been studied and
an application of these results to HST data is in progress (Mayen C. \&
Soucail G., in preparation). 

\acknowledgments

We wish to thank the European TMR Programme \textit{Gravitational
Lensing : New Constraints on Cosmology and The Distribution of Dark
Matter} for its financial support.


\begin{references}
\reference Broadhurst, T.J., Taylor, A.N., Peacock, J.A., 1995, \apj,
438, 49
\reference Fort, B., Mellier, Y., Dantel-Fort, M., 1997, \aap, 321, 353
\reference Mayen, C., Soucail, G., 1999, \aap, in preparation
\reference Schneider, P., Ehlers, J., Falco, E.E., 1992, Gravitational
Lenses, Berlin
\reference Taylor et al., 1998, \apj, 501, 539
\end{references}
\end{document}